\documentstyle[aps, pra, tighten, preprint]{revtex}
\begin{document}
%
%
\hoffset= -2.5mm

\title{Mesoscopic Noise Theory: Microscopics, or Phenomenology?\cite{upon} }

\author{F. Green\cite{fgemail} \\}
\address{
GaAs IC Prototyping Facility,
CSIRO Telecommunications and Industrial Physics,
PO Box 76, Epping NSW 1710, Australia \\}

\author{M. P. Das\cite{mdemail} \\}
\address{
Department of Theoretical Physics,
Research School of Physical Sciences and Engineering, \\
The Australian National University,
Canberra ACT 0200, Australia \\}

\maketitle

\bigskip

\begin{abstract}
We argue, physically and formally, that existing diffusive models of noise
do not yield a faithful microscopic description of nonequilibrum current
fluctuations. The theoretical deficit becomes more evident in quantum-confined
metallic systems, such as the two-dimensional electron gas. In such systems
we propose a specific experimental test of mesoscopic validity for
diffusive theory's central claim:
the smooth crossover between thermally induced and shot-noise fluctuations
of the current.

\end{abstract}

\bigskip
\bigskip


\section*{Introduction}

Modern developments in noise physics,
theoretical and experimental,\cite{kogan,djb}
have opened a remarkable window on charge transport and fluctuations
in mesoscopic systems.
At scales of tens of nanometers or less,
comparable to mean free paths for scattering,
it is hard to justify the usual simplifications of
long-range homogeneity that characterize transport in the bulk.
Moreover, such small systems are wholly open to the larger environment
and experience substantial, rapid exchanges of energy and particles.
Unavoidably, mesoscopic conductors undergo
large excursions from equilibrium.
For instance, in a good field-effect transistor a
source-drain potential of just 0.1 V, across a gate region
100 nm long, sets up a mean gradient of $10^4 {\rm ~V ~cm}^{-1}$.
Typical driving fields are no longer weak.

Linear-response theory then becomes relatively narrow in scope.
Despite this, most work on mesoscopic metallic  conduction
favors the low-field, near-equilibrium properties
of transport and noise, with less activity in the
technologically relevant nonequilibrium regime.
(An exception to the general
neglect of degenerate high-field fluctuations is the Monte Carlo study
by P. Tadyszak {\em et al.}.\cite{mc})
Of course, within the low-field domain there
have been striking successes.\cite{djb} Best-known
is the prediction\cite{beebut,ml,nagaev}
and observation\cite{liefr,steinbach}
of the threefold suppression of Poissonian shot noise. This occurs
via Pauli blocking of coherent electron correlations, when
transport is dominated by elastic scattering in a metallic wire.

Other mesoscopic shot-noise effects of potential
importance have now come to light. One thinks of
threefold space-charge suppression, predicted for
nonuniform systems of classical carriers.
\cite{regg}
This is an example of an inherently high-field process
where linear methods fail.
In the critique below, we recall some unresolved issues in the noise
theory of metallic conductors, with an eye to
strongly nonequilibrium situations. That is where practical
mesoscopic devices will normally operate, and where the standard,
strictly linear, formalisms \cite{kogan,djb,datta}
lose much of their relevance.

A theoretical program for mesoscopic fluctuations, covering
both high- and low-field cases from the start,
will constrain its physics tightly
in both limits and thus everywhere between.
By comparison, in predicting an effect like the
{\em smooth transition between
thermal current noise and shot noise},\cite{djb,beebut,ml,nagaev}
existing theories transcend their own, self-imposed, linearity.
\cite{datta}
Other theories, set up to be nonperturbative in the driving field,
may account very differently for such nonequilibrium effects.

Well-controlled high-field descriptions of mesoscopic
fluctuations must meet two criteria. (a) They should
be microscopic, in the sense of kinetic theory.\cite{kogan}
For example, a
semiclassical formalism should add as few assumptions as possible
to those already built into the Boltzmann equation. (b)
They should go to equilibrium naturally and seamlessly.
For example, the fluctuation-dissipation relation\cite{kogan}
must always follow from the models' axioms.

In the Section immediately following, we recall the
leading approaches to mesoscopic fluctuations with reference
to principles (a) and (b).
Next, we outline a nonperturbative semiclassical
model that satisfies (a) and (b) and deals with
high- and low-field fluctuations equally.\cite{gd}
Then we address the thermal- to shot-noise crossover;
we propose basic measurements to resolve the mutually exclusive
predictions of linear versus nonperturbative theories,
with an outline of the implications of an experimental disproof
of the smooth crossover. We end with a summary.

\section*{Low-Field Linear Theories}

Theories of mesoscopic noise are either semiclassical, based on
the Boltzmann transport equation for degenerate electrons,
\cite{djb,nagaev}
or quantum-mechanical, based on the Landauer-B\"uttiker picture of conduction
as a unitary {\em S}-matrix process.\cite{djb,beebut,ml}
Although the two approaches are calculationally very different,
their noise predictions are frequently almost identical.
This comes about through a common handling of
the boundary-condition problem at the conductor's
interfaces with the environment.
Both philosophies reach near-identical conclusions
because they adopt identical premises for the reservoirs.
\cite{datta}

Before reviewing the diffusive
viewpoint, credited to Landauer,\cite{ldr57,ldr88}
we note that the issue of attaching a conductor
to its external leads is not fully resolved.\cite{henny}
The conceptual problems of connecting to
carrier states in the reservoirs become acute
as soon as one leaves the familiar linear limit.
Among others we refer to the work of Frensley,
\cite{fren} Johnson and Heinonen,\cite{jh}
and the novel quantum analysis of Magnus and Schoenmaker,
who address energetics and gauge invariance.
\cite{ms}

\subsection*{Drift-Diffusion}

Diffusive noise analysis works with an Ansatz
specific to {\em low-field} transport. This
replaces the kinetic description of a drift flux in a local driving
field with the description of a diffusive flux
along an effective local density gradient.\cite{datta}
The approach was inspired by Landauer's
treatment of localized scattering within a metallic host.
\cite{ldr57,ldr88}

In all such models, microscopic field distributions
no longer appear. Only the overall electromotive potential
is referenced directly. In particular, drift-diffusion equivalence
leads to a description of the (equilibrium) carriers at the
source and drain, couched
solely in terms of the mismatch in their local Fermi levels
relative to some effective global band edge.
This level difference naturally accompanies the difference
in the effective density of carriers diffusing across the channel,
and becomes identified with the electromotive potential.
The rationale for determining a {\em global} band-edge zero
is not discussed much.
\cite{kogan,djb,beebut,ml,nagaev,datta}
Elsewhere, however, the device literature is forced to wrestle
with the tough problem of accurately locating the {\em physical} band
edge, which varies all along the extended channel.
\cite{selb}

For conductance investigations, the drift-diffusion
replacement is justified on two phenomenological grounds.\cite{datta}
First, if a small driving potential $eV \ll \mu$ is applied,
where $\mu$ is the Fermi level, the response must surely be
linear. The Einstein relation should hold overall,\cite{datta}
giving a net low-field conductivity $\sigma$ that is
completely interchangeable with the diffusion constant $D$:
\cite{ldr57,vvt}

\begin{equation}
\sigma {~\longleftrightarrow~} e^2 D {\partial n\over \partial \mu},
\label{eq1}
\end{equation}

\noindent
where $n$ is the electron density.
Second, Pauli degeneracy within a dense population of
electrons permits scattering transitions only in a small band
around the Fermi energy,
of the order of the thermal energy $k_BT \ll \mu$. All other carriers
cannot scatter because of Pauli blocking; this
non-participating background drops out
of the calculation of transport.

\subsection*{Fluctuation-Dissipation}

The successful solution of many conductance problems
invites firm confidence in the power of diffusive analysis.\cite{imry}
What, then, could hinder its applicability to fluctuations?
The answer depends crucially on the
hypothesis of drift-diffusion equivalence.
Equation (\ref{eq1}) relies on the fluctuation-dissipation theorem
(FDT), in its Einstein formulation. The theorem,
valid within linear response, establishes
a unique correspondence between dissipation due to the mean current
($\sigma$ is determined by one-body scattering) and mean-square fluctuations
away from the zero-current equilibrium state
($D$ is determined by two-body correlations, induced by scattering).
The FDT is merely one of several essential relations that link the
single-particle and two-particle density matrices in a fundamental way.
Chief of these are the Ward identities\cite{nozi} and their offshoots,
the compressibility and perfect-screening sum rules.
\cite{nozpin}

Diffusive transport models obtain the mesoscopic conductance
from a hydrodynamic drift-diffusion Ansatz.
The justification for replacing drift with
diffusion is the FDT. This theorem
can be derived {\em only} within a microscopic description.

Short of supplying a microscopic proof of the Einstein
relation {\em for mesoscopic systems},
by the Kubo formula\cite{ziman,sols} or otherwise,
diffusive models have only the bulk form of the relation to draw upon,
with all the qualifications that hedge its adoption for mesoscopic
problems.\cite{ldr88}
No matter how reasonable it is to presume linearity,
the validity of drift-diffusion equivalence
remains hypothetical within such treatments.
This creates at least three unsolved problems:

\begin{itemize}
\item The status (and meaning) of the FDT
is an open question for such models since, if the result is assumed,
it cannot be derived. If it cannot be derived, neither is
it possible to access its core ingredient: the fluctuations.

\item Invoking the FDT
enforces a strict, and inescapable, linear structure upon all the results
of the models. Out of the linear regime, which is {\em not} where
practical mesoscopic devices will operate, that strategy becomes unusable.

\item The added requirement of strong degeneracy
limits the description
to low temperatures. One cannot handle intermediate,
moderately degenerate cases where $\mu \lesssim k_BT$, since then
almost all states are partially occupied and one has to fall back
on kinetic methods. The moderately degenerate electron gas is ubiquitous:
it characterizes most two-dimensional microelectronic devices under
normal operating conditions.
\end{itemize}

\subsection*{Compressibility, Closure, and Completeness}

We come to the diffusion-based theories' ability to
capture fluctuations and noise, in view of
the incompressible nature of the electron gas.
Metallic-electron physics is overwhelmingly governed by
degeneracy and strong Coulomb screening.\cite{nozpin}
This is the case at all scales longer than the de Broglie wavelength,
including mesoscopic distances. Mesoscopically, the dominant feature
is still the {\em incompressibility} of the underlying Fermi sea.
Compressibility is basically a two-body effect.
\cite{nozpin}
Indeed, it is the extraordinary stiffness of the electron gas
that conditions its correlations.

The importance of treating the incompressible Fermi sea as a unity
is a challenge for diffusive models,
which describe -- by definition -- a {\em compressible}
medium along the mesoscopic wire,
made up only of those carriers that partially fill
states near the Fermi level.\cite{datta}
Filled states in the Fermi sea are discarded
and play no role thereafter, even
though it is clear from the theory of metals
that all the electron states mediate
degeneracy and mean-field screening, and that all must
enter into the collective behavior of the fluctuations.
A diffusive model is not designed to establish the Einstein relation
or the FDT, and it is not designed to uphold the
compressibility and perfect-screening sum rules.\cite{nozpin}

Finally we examine the need for a constitutive relationship
(closure) for the electron fluctuations
in terms of, say, the current through the system.
There is no analogous ``super-Einstein'' relation to direct
the diffusive evolution of the fluctuations; there is
only the knowledge encoded, and already exploited, in Eq. (\ref{eq1}).
Here, it is important to recall that the fluctuation physics
is actually {\em internal} to the coefficient $D$,
a velocity autocorrelation.\cite{vvt,fg}
By construction, diffusive phenomenology shuts
itself off from analytical access to the microscopic structure of
the diffusion constant.

If $D$ has the status of a primitive parameter in Eq. (\ref{eq1}),
as it must within diffusive phenomenology,
that status divorces the diffusion constant from its
key sensitivity to the underlying fluctuation physics
(see the following Section).
One is left with just two alternatives for a closure
relying on Eq. (\ref{eq1}). Either the fluctuations
of diffusive noise theory are hermetic,
relating solely to themselves in a trivially circular way,
or else one is forced to guess them as functions of the only
object to which Eq. (\ref{eq1}) gives open access:
the mean single-particle occupancy.

To make a guess about fluctuations is again to beg a
question. In the language of statistical mechanics:
Is the two-body density matrix of a quantum ensemble
canonically representable in terms of the diagonal part
of the one-body density matrix? This is a risky thesis
because it means that the complete set of excited states for
the system is truncated arbitrarily.\cite{rmrk}

In linear response it is completeness of the microstates that
determines every statistical average and underlies the sum rules.
The exponential Boltzmann weighting for any given multipair state
may be insignificant, but combinatorially such excitations are myriad.
None of the diffusive models has a systematic way to
classify them and cull them (as, for example, the classic
random-phase and ladder approximations do).
For that, a hierarchy of kinetic equations is essential.\cite{ziman}

Little can be done, after the fact, to reinforce
the diffusive Ansatz with kineticlike arguments.
\cite{kogan,djb,beebut,ml,nagaev}
They are not enough to undo its phenomenology, nor to remedy
its structural incapacity to describe the correlations.
By all means, such arguments convey seminal physical ideas,
as shown by threefold suppression of
shot noise in elastically dominated systems. Neverthless,
for now, they have no guarantee of compatibility with
collective properties of the degenerate electron gas.
\cite{nozpin}

Our point is not at all to query the existence of
a mesoscopic fluctuation-dissipation theorem.
On the contrary: to know the systematic fluctuation
properties of electronic systems -- small and big -- one must 
know the form of their FDT rather than hypothesize it.
That form is manifest only through
explicit derivation within statistical mechanics
or kinetic theory. This has always been the case,
and it remains the case regardless of physical scale.

In the next Section we indicate that
a canonical and {\em computable}
kinetic theory of mesoscopic fluctuations is feasible.
Such a description can be (and should be)  set up as an inherently
{\em nonequilibrium} microscopic model. Necessarily, it
treats the {\em whole} ensemble of degenerate carrier states
and thus preserves the dominant physics of the Fermi sea.
Last, it generates a proof of the mesoscopic FDT.

\section*{A High-Field Kinetic Theory}

Reference \onlinecite{gd} sets out the complete mathematical specification
for a standard kinetic theory of mesoscopic noise,
with the following characteristics.

\begin{itemize}
\item {\em Formal structure}.
The theory relies on Green-function solutions to the
linearized Boltzmann equation. This gives the semiclassical
form of the particle-hole correlations in an electron gas driven
out of equilibrium. The model is Markovian; as such, its
equivalence to the Boltzmann-Langevin formalism is often stated.
\cite{kogan}
Its solutions are {\em nonperturbative} in the applied field,
hence not bound to a low-order expansion restricting applicability.

\item {\em Explicit form of nonequilibrium noise}.
The theory solves the exact nonequilibrium distribution
of electron-hole pair correlations as a calculable, linear functional
of its known equilibrium form.\cite{nozpin}
In degenerate systems this leads to a core result:
thermal fluctuations, including the excess hot-electron terms,
are {\em always} proportional overall to the temperature $T$ of
the ideal thermal bath.
Shot noise does not scale with $T$ and thus has
{\em no link} to excess thermal noise.

\item {\em Noise suppression by self-screening}.
A major feature is Coulomb self-screening, suppressing the current
fluctuations whenever a {\em contact potential} characterizes the system
interface with its ideal electron reservoir.
Suppression is defined by a factor $\gamma_C(n)$ that
varies sensitively with electron density,
particularly in two-dimensional quantum-well channels.\cite{2deg}
As with their $T$-scaling,
both equilibrium {\em and} nonequilibrium charge fluctuations
must scale with $\gamma_C(n)$, the latter
doing so because they depend directly on the former.
There are implications for low-noise heterojunction
devices.\cite{gc}

\item {\em Fluctuation-dissipation theorem}.  
Away from equilibrium the theory quantifies the connection
between the current-fluctuation spectrum and dissipation,
or Joule-heating rate.
Near equilibrium, that connection becomes the microscopic FDT.
While for quantized Coulomb systems there is
self-induced suppression of the fluctuations,
there can be no violation of the thermodynamic
noise-power balance at the heart of the FDT.
\cite{kittel}
Rather, the kinetic part of the internal energy of the
fluctuations is now complemented by a large, self-consistent
potential part; these are not separable in Maxwell's
sense, but the spectral density of their sum
still obeys the Johnson-Nyquist relation.
The {\em purely kinetic term}, and the closely related
{\em velocity-velocity correlation} along
with its {\em effective noise temperature}
(the last two implicit in the Einstein relation\cite{fg}),
are all renormalized by $\gamma_C$. This heralds new effects.
\end{itemize}

The semiclassical analysis rests squarely on two pillars:
Boltzmann kinetics\cite{kogan} and Fermi-liquid theory.\cite{nozpin}
We stress its utter conformity with standard device physics,
\cite{selb}
both as to assumptions and boundary conditions.
The theory must have lower bounds on the
length and time scales for reliable noise prediction.
However, by incorporating the
complete fluctuation structure of the electron gas,
whose screening length and plasmon period are extremely short,
proper mesoscopic calculations lie well within its scope.

Fully kinetic approaches of this kind are not phenomenologically
tied to linear response and do not merely ape the output of
Boltzmann-Langevin analysis (certainly not of its diffusive variants).
\cite{kogan}
They go much further towards addressing strongly nonequilibrium
situations. These are completely intractable within Boltzmann-Langevin
and, for that matter, Landauer-B\"uttiker models of noise.

A fruitful and testing area of study is shot noise.
We now survey a nonperturbative approach to shot noise via
truly kinetic methods, stressing differences with
phenomenological approaches and their predictions.
Space prevents a full account here; we focus on physical implications,
offering Ref. \onlinecite{gd} as a technical resource.

\section*{Shot Noise and The Crossover Effect}

\subsection*{Microscopic Description of Shot Noise}

The description of mesoscopic shot noise should have an operational
thrust, reflecting the affinity with time-of-flight experiments:
at some time an electron is launched from the cathode,
and later detected at the anode.
This differs from prescriptions in which fluctuations of the
circuit current are volume-averaged to make use of the
Ramo-Shockley theorem, then autocorrelated.
\cite{kogan,djb,beebut,ml,nagaev}
Such a procedure reflects the physics of distributed thermal noise,
not of particulate noise.

A two-terminal mesoscopic conductor is very similar to the region
enclosed between the cathode and anode of a traditional vacuum tube.
Shot noise consists of a current fluctuation at one terminal
correlating with a current fluctuation at the other.
While thermal fluctuations directly probe the whole volume of a conductor,
shot noise {\em indirectly} probes the charge dynamics internal to
the enclosed region. Both coexist, and are stochastically independent.

In a classical macroscopic circuit, the time-of-flight
viewpoint adds nothing to that of spatial averaging
because, in that case, the two are ergodically equivalent.
However, in degenerate mesoscale conductors,
the two descriptions are not the same and
shot noise emerges as a nonlocal, two-point correlation.
These ideas have a direct thermodynamic basis, as we show below.

An elementary shot-noise event adds or removes precisely
one carrier in the system, randomly in time. Temporal randomness
makes the total signal the incoherent sum of elementary events.
Each of the $N$ carriers dwelling in the sample contributes
equally to the stochastic sum; thus the shot noise across its
two terminals becomes

\begin{equation}
S_{\rm shot}(c;a) =
2 N {\left\{ |\Delta N|
{ {\langle {(-ev\delta f_x) \star (-ev'\delta f_{x'})} \rangle}
\over {\delta N} }\!{}_{x \in c; {~}x' \in a}
\right\}}.
\label{eq2}
\end{equation}

\noindent
This correlates the notional change
$-ev \delta f_x$ in flux distribution at the cathode ``$c$'' with the
change $-ev' \delta f_{x'}$ at the anode ``$a$'', both induced by a
{\em discrete} change $\Delta N = \pm 1$ in $N$.
The notation $\langle ~ \star~ \rangle$ denotes the trace
of the correlated (electron-hole)
fluctuation over {\em all} the microscopic particle states
in the cathode and anode regions of the conducting system, and
the denominator $\delta N$ normalizes the correlated response
to a unit fluctuation of the total electron number.
Eq. (\ref{eq2}) recovers all the familiar results for simple shot noise.
For example, it yields the usual Schottky result
$S_{\rm shot} = 2eI$ for a current $I$ of quasi-monoenergetic
carrier wavepackets, with Poissonian arrival statistics.
Its {\em high-field} properties are more engaging.
\cite{gd}

The correlation dynamics for Eq. (\ref{eq2}), as moderated by
scattering, are given by the {\em Green function} for the equation
of motion.\cite{gd}
This function is contained within the object
$\delta f_x \star \delta f_{x'}$, whose physics is that of
a correlated electron-hole propagator (its form is de-emphasized
here, purely to focus on scaling properties of the shot noise).
Ready access to the propagators means that this kinetic noise theory
remains adaptable and practical, even far from equilibrium.

\subsection*{Scaling}

Rewrite Eq. (\ref{eq2}) as

\begin{eqnarray}
S_{\rm shot}(c;a)
=
&&
2 e^2
N { {\langle vv' {\delta f_x \star \delta f_{x'}} \rangle}\over \delta N}
\!{}_{x \in c; {~}x' \in a}
\cr
{\left. \right.} \cr
=
&&
2 e^2
N { {k_BT [{\langle vv' {\delta f_x \star \delta f_{x'}} \rangle}/
\delta \mu]}\over {k_BT [\delta N/\delta \mu]} }
\!{}_{x \in c; {~}x' \in a}.
\label{eq3}
\end{eqnarray}

\noindent
Both numerator and denominator are rescaled homogeneously,
each becoming explicitly proportional to the
thermal fluctuations in the structure.
That step makes sense {\em if and only if} the nonequilibrium
correlations are unique linear functionals of the equilibrium ones;
the key property of this model.
Equilibrium fluctuations are given by
$k_BT [\partial f^{\rm eq}/\partial \mu$] where $f^{\rm eq}$ is
the single-particle distribution.

If the fluctuations undergo significant Coulomb self-screening,
the additional suppression factor $\gamma_C(n)$ appears with $k_BT$.
The thermal fluctuations of current density, integrated over
volume $\Omega$ of a conductor with length $L$, correspond to

\begin{equation}
S_{JJ}(\Omega)
= 4 \gamma_C(n) k_BT
{\left\langle
{\Biggl( {-ev\over L} \Biggr)} {\Biggl( {-ev'\over L} \Biggr)}
{{\delta f_x \star \delta f_{x'}}\over \delta \mu}
\right\rangle}_{x, x' \in {~\Omega}}
\label{eq4}
\end{equation}

\noindent
(in which $[\delta f_x \star \delta f_{x'}]/\delta \mu$
is to be computed in the absence of self-screening).
In the uniform limit, where $\gamma_C(n) \equiv 1$, this expression
is the spectral distribution of Johnson-Nyquist noise.
For strongly driven, nonuniform, or quantum-confined systems
it predicts a rich variety of thermal fluctuation behaviors.
\cite{gd}
In a partially self-confined electron gas
at a heterojunction, the current fluctuations
should be suppressed:
$S_{JJ} \to 4Gk_BT_{\rm eff}$, where $G$ is the
standard conductance and $T_{\rm eff} = \gamma_C(n)T$
is the effective temperature
\cite{gc,kittel}
of the fluctuations in current density.

Note that $S_{JJ}$ then no longer represents
total thermal noise, but its kinetic component alone
(recall the previous Section). Therefore direct
experimental access to $S_{JJ}$ must go beyond
the means normally adequate for Johnson-Nyquist noise.

Comparing Eqs. (\ref{eq3}) and (\ref{eq4}), it is immediate that
the shot noise cannot scale with $T_{\rm eff}$.
The underlying physics is simple.
Shot noise is sensitive, first and last,
to the discrete addition or subtraction of single carriers.
In this, the environmental conditions of temperature,
and of the electrostatics of the interfaces, play no role.
Shot noise will not feel those effects explicitly.

\subsection*{Noise and Thermodynamics}

The form of Eq. (\ref{eq3}) is not an academic exercise.
It is a physical necessity.
Consider the $T$-dependence of thermal fluctuations in a
degenerate system. Electron correlations that are thermally
driven must always
scale with ambient temperature, since $T$-scaling is {\em mandatory}
for the exact solution to the two-body kinetic equation.
\cite{gd}
This is because
(i) the nonequilibrium correlations stand in precise linear
relation to the equilibrium ones, and (ii) the fluctuation-dissipation
theorem uniquely imparts Johnson-Nyquist normalization to the thermal
correlations, in the low-field limit.
If there are contact-potential effects in the system,
item (i) clearly implies an additional Coulomb-induced rescaling
of the correlations through $\gamma_C(n)$.

Proportionality to $T$, at any driving field,
is a strict requirement for degenerate thermal fluctuations;
\cite{lagen}
its outworking is Eq. (\ref{eq4}).
That is why it is not possible to treat shot noise in a metallic conductor
on the same footing as its thermal noise. Shot noise does
not scale with temperature (there is agreement
on this at least). From a thermodynamic perspective,
shot noise cannot be described at the same formal level as thermal noise,
for it has a distinct physical nature.

It is then straightforward that shot noise arises precisely from
fluctuations of the number of carriers in transit through the conductor.
Variations with respect to particle number are nothing
other than the thermodynamic conjugates of variations with respect
to the electro-chemical potential, describing the thermal
current fluctuations.
Though intimately related microscopically,
the two effects could hardly be more different quantitatively.
\cite{td}

The difference in scale between shot-noise and thermal
current fluctuations is fundamental and irreducible.
Any continuous transformation linking one
to the other is therefore inadmissible
(except classically, where they merge\cite{class}).
Such behavior contrasts sharply with the prediction
of diffusive models.
There, a {\em continuous and universal} transition
is obtained, between thermally driven and shot-noise
fluctuations, as the applied potential exceeds $2k_BT$:
the smooth crossover.
\cite{kogan,djb,beebut,ml,nagaev}

\subsection*{A Test of the Smooth Crossover}

We end by proposing a current-fluctuation measurement to
distinguish between the diametrically opposed predictions.
Consider a two-dimensional electron gas, typically formed at a
III-V heterojunction. The
high-band alloy, usually heavily Si-doped, should be doped
lightly to moderately for enhancement-mode operation.
The channel density should be controlled by back-gate biasing from
the substrate side of the device, for uniform control over
the whole area of the structure. On the high-band
alloy side one might mesa-etch a mesoscopic wire, or else
deposit a split gate to define the wire electrostatically.
Finally, make ohmic contacts to the wide access regions of the channel, 
feeding the wire proper. The operating width of the wire
should be generous, not less than 0.5 $\mu$m,
to simplify theoretical analysis.

Altering the channel density by the back-gate voltage
should alter the effective temperature $T_{\rm eff} = \gamma_C(n)T$,
a sensitive function of $n$.
Reduction of the free-carrier fluctuations can be substantial;\cite{gc}
for $n \sim 10^{12}~{\rm cm}^{-2}$, usual in
AlGaAs/InGaAs/GaAs quantum wells, one has $\gamma_C(n) \approx 0.45$.
When $n \lesssim 10^{11}~{\rm cm}^{-2}$, then $\gamma_C(n) \gtrsim 0.8$.
The thermally driven current correlations (though not their
{\em total} energy density) should directly
mirror the strong bias dependence of $\gamma_C(n)$.

If, as we predict, shot noise does not depend on $\gamma_C(n)$
but primarily on source-drain current -- with {\em or} without other,
physically unrelated mechanisms for suppression 
\cite{djb,regg,gd} -- its contribution to the current fluctuation
spectrum will change far less with bias, for any current level.
Resolving the direct spectrum into its incommensurate parts,
$S_{JJ}$ and $S_{\rm shot}$, should not be too hard
with a good characterization
of the quantum-well structure, and a good calculation
of $\gamma_C(n)$. The latter has long been a well-understood
element of device-gain modeling.\cite{gc}

The persistent mismatch between the two correlation scales, thermal
(scaling with $\gamma_C$) and shot (independent of $\gamma_C$),
is inconceivable within linear diffusive theories. Instead they predict
a continuum of hybrid current fluctuations,
interpolating between the free-carrier Johnson-Nyquist
result (with no $\gamma_C$) and shot noise,
and parametrized by the applied voltage in a most nonlinear way.
\cite{djb,beebut,ml,nagaev}
Experimental disproof of the smooth crossover would
call for revision of such models, and a fresh
understanding of data apparently confirming the crossover.
\cite{steinbach}

Not least, this experiment promises a
{\em direct physical measurement} of $\gamma_C(n)$.
Self-screening of current fluctuations in the
quantum-well channel
may be a major reason for the excellent low-noise behavior of
heterojunction field-effect transistors,\cite{gc}
a vital microelectronic technology.
Since, in any case, $\gamma_C(n)$ influences the
intrinsic gain and capacitance of such structures,\cite{gc}
noise-based studies of the physics of $\gamma_C(n)$ may
offer a new diagnostic tool
for high-performance millimeter-wave designs.

\section*{Summary with Open Question}

Diffusive mesoscopic theories rely on an
efficient phenomenology for conductance, but
for noise investigations the same approach is questionable.
Assuming a mesoscopic Einstein relation,
rather than deriving it microscopically, ties these 
models to strict linearity. They do not suit the genuinely
nonequilibrium transport and noise problems that are the
rule, not the exception, in nanodevice engineering.

At metallic densities, the insistence on a freely compressible
methodology for electron fluctuations is problematic.
Degenerate fluctuations are dominated by the stiffness
of the electron gas. Thus, to ignore the rigid constraints on
compressibility
\cite{nozpin}
is to risk misrepresenting the main effect of degeneracy.
We have also argued that diffusive theory has no logical
connection to orthodox kinetics, leading to non-compliance
of its correlations with the norms of statistical mechanics.

Alongside the above goes the failure
to allow for large quantum departures
from classical equipartition of energy. This means that
the energetics of strongly confined Coulomb systems
are misrepresented, right at the microscopic level.
Such systems form the substrate
on which a large part of mesoscopic technology depends.

It is feasible to set up a {\em conventional}, and calculable,
semiclassical kinetic theory of mesoscopic carrier fluctuations.
Its consistent, nonequilibrium microscopic description conforms
to statistical mechanics and obeys the sum rules. This allows
a rigorous derivation of the FDT, including the case of
quantum-confined metallic systems.
Such a description entails a clear-cut separation of scales
between shot-noise and thermal correlations
of the current in degenerate conductors.
It negates any possibility of transmuting
one correlation effect into the other, continuously.

Kinetic theory leads to the incommensurability of
thermal and shot-noise fluctuations.
This result, untrammeled and unambiguous,
suggests that it may yet be too soon to canonize the
smooth crossover.
The relative scarcity of direct data
on current-current correlations in low-dimensional metals,
\cite{lagen}
even for plain thermal effects,
invites our proposal for a sharper experimental test
of the smooth crossover.

In the end there is but one open question:
What will experiment say?

\section*{Acknowledgments}

We thank Prof. Lino Reggiani for his robust and
constructive criticism of our ideas, helping us
to refine our feeling for the issues.
We also record our debt to Rolf Landauer.
Though in deepest disagreement with our end results, he
gave us crucial encouragement to enter the field.



\begin{references}
\bibitem{rmrk}
A very clear assertion of the importance of off-diagonal
contributions, even for purely one-body processes, is
made by Landauer, Ref.\onlinecite{ldr57}.
\bibitem{2deg}
In a quantum-confined two-dimensional electron gas, $\gamma_C(n)$
is the thermally induced fluctuation of kinetic energy
at the Fermi surface, as a fraction of the fluctuation
in total energy (kinetic and electrostatic sum to $k_BT$).
Classical equipartition precludes such suppression;
it is a purely quantum negative-feedback effect.
See Ref. \onlinecite{gc}.
\end{references}
\end{document}